\documentclass[aps,prd,twocolumn,groupedaddress]{revtex4-2}
\usepackage{graphicx}
\usepackage{dcolumn}
\usepackage{bm}
\usepackage{xspace}
\usepackage[colorlinks=true, allcolors=black]{hyperref}
\def\v4641{V4641 Sgr}
\def\gr{$\gamma$-ray}
\def\aap{A\&A}
\def\apjl{Ap.J.Lett.}

\begin{document}

\title{Multi-messenger signature of cosmic rays from the microquasar \v4641\ propagating along a Galactic magnetic field line}
\author{A.~Neronov$^{1,2}$, F.~Oikonomou$^{3}$  and D.~Semikoz$^{1}$}
\email[]{andrii.neronov@apc.in2p3.fr,foteini.oikonomou@ntnu.no,\\dmitri.semikoz@apc.univ-paris7.fr}
\affiliation{$^1$Université de Paris Cite, CNRS, Astroparticule et Cosmologie, F-75013 Paris, France}
\affiliation{$^2$Laboratory of Astrophysics, Ecole Polytechnique Federale de Lausanne, CH-1015, Lausanne, Switzerland}
\affiliation{$^3$Institutt for fysikk, NTNU, Trondheim, Norway}

\begin{abstract}
The recently detected extended very-high-energy \gr\ emission from the microquasar \v4641 reveals a puzzling 200-parsec-long jet-like structure significantly misaligned with its radio jet. We propose that this \gr\ structure is produced by high-energy cosmic-ray particles escaping from the microquasar along ordered field lines of the Galactic Magnetic Field and interacting with the interstellar medium. 
If the \gr\ emission is produced by interactions of high-energy cosmic ray nuclei, the system is detectable by future multi-km3 neutrino detectors. We argue that \gr\ observations of jet-like features adjacent to high-energy sources in the Milky Way provide a new method to measure the regular and turbulent components of the Galactic magnetic field at different locations in the Milky Way. 
\end{abstract}

\newcommand{\Sgr}{V4641~Sgr\xspace}
\newcommand{\km}{KM3NeT\xspace}
\newcommand{\ic}{IceCube\xspace}
\newcommand{\gvd}{Baikal-GVD\xspace}
\newcommand{\gray}{$\gamma$-ray\xspace}
\newcommand{\grays}{$\gamma$-rays\xspace}

\maketitle


\section{Introduction \label{intro}}

Our knowledge of the Galactic Magnetic Field (GMF) mainly stems from radio observations of synchrotron emission from the interstellar medium, Faraday rotation measurements, and polarized dust emission \cite{1996ARA&A..34..155B}. In many cases, the measurements only constrain the integral characteristics of the magnetic field, such as the line-of-sight integral of the parallel components of the field strength weighted with the free electron density in the case of the Faraday rotation technique and the line-of-sight integral of the perpendicular component of the field weighted by the relativistic electron density in the case of synchrotron emission. This leads to significant uncertainties in the knowledge of the field structure \cite{2012ApJ...757...14J,Unger:2023lob,Korochkin:2024yit}. 

The Galactic magnetic field structure determines the propagation regime of high-energy particles in the interstellar medium. Particles can stream along ordered magnetic field lines and scatter on inhomogeneities introduced by the turbulent field. Depending on the relative strength of the ordered and turbulent field components, a combination of streaming and scattering is expected to lead to more or less isotropic diffusion of particles through the medium \cite{Giacinti:2012ar,Giacinti:2013wbp,Giacinti:2017dgt}. 

High-energy particles propagating through the interstellar medium interact and produce emission detectable with \gr\ telescopes. This emission can potentially be used to probe both the GMF and the details of particle propagation. An example is given by pulsar halos in which inverse Compton scattering by high-energy electrons and positrons released by a pulsar wind nebula is observed in the TeV energy range \cite{2017Sci...358..911A}. The observational appearance of pulsar halos suggests that particles spread nearly isotropically within tens of parsecs around the source and that the diffusion coefficient is much smaller than expected based on modelling the escape of cosmic rays from the Galaxy in the conventional isotropic diffusion model framework \cite{1990acr..book.....B}. It is not clear if this points to a peculiarity of particle diffusion or of the magnetic field structure in the interstellar medium in the vicinity of a specific high-energy particle source class (the middle-aged pulsars) around which the halos are found  \cite{Schroer:2023aoh}. To get more precise information on the propagation of cosmic rays through the GMF, one must find other examples of high-energy particles spreading away from high-energy sources.  The extended sources of the highest energy \gr s discovered by  HAWC~\cite{2020ApJ...905...76A} and LHAASO~\cite{LHAASO:2023rpg}  are the best candidates for exploring this possibility. 

The candidate sources for the study of diffusion through the GMF are not necessarily expected to be similar to the pulsar halos. If the turbulent field dominates, the diffusion is expected to be isotropic on distance scales larger than the field correlation length, but subtle anisotropic patterns may emerge on shorter distance scales \cite{Giacinti:2013wbp,Bao:2024rrg,Bao:2024cdv}. When the ordered field is stronger than the turbulent field, particles spreading from their injection point preferentially spread along the ordered field lines and the diffusion may remain strongly anisotropic on larger scales \cite{Giacinti:2022ohy}.

In what follows, we discuss the possibility that the recently discovered new class of the highest energy \gr\ sources, microquasars,~\cite{lhaaso} may provide a probe of the GMF and of high-energy particle diffusion through the interstellar medium not perturbed by recent star formation activity associated to supernovae and superbubbles.  Microquasars, and especially microquasars in low-mass X-ray binaries, like GRS 1915+105, \v4641 V404 Cygni \cite{lhaaso,hawc,MAGIC:2017oci}, are ``old" systems that had enough time to move away from their star formation regions and hence a study of particles escaping from these sources provides a complementary view of particle diffusion through the GMF, compared to pulsar halos. Some of these sources are surrounded by extended emission that may be similar in nature to the pulsar halos. We specifically consider the example of the \v4641 system and its almost two-degree-scale extension \cite{hawc,lhaaso} for which we propose an interpretation in terms of emission from particles diffusing away from the source. 

\section{\v4641}

\begin{figure}
    \includegraphics[width=\columnwidth]{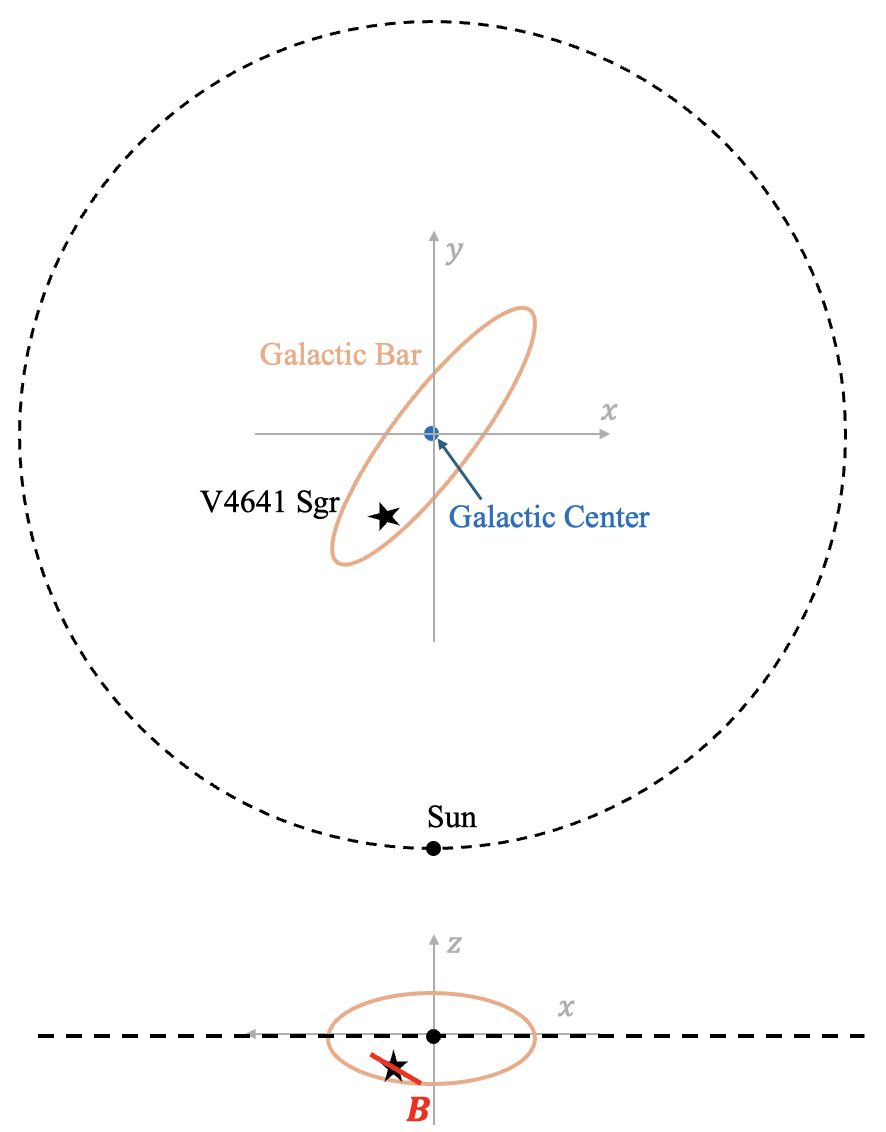}
    \caption{Location of \v4641\ in the Milky Way with respect to the Galactic Bar 
    \cite{2019A&A...628A..94A}. The red line shows the inferred direction of the GMF projected on the $yz$ plane. Top and bottom panels show $xy$ and $xz$ projections of the Milky Way, with the rotation axis of the Galactic disk aligned along $z$ axis. The dashed line shows the solar circle. }
    \label{fig:cartoon}
\end{figure}

\v4641, recently detected at energies reaching beyond 100~TeV by LHAASO \cite{lhaaso} and HAWC \cite{hawc}, is a microquasar at a distance $6.6$~kpc \cite{Gaia:2018ydn} hosting a black hole of mass $M_{bh}\simeq 6.4M_\odot$ orbiting a $\simeq 3M_\odot$ companion star on a $2.8$~day orbit \cite{MacDonald:2014gpa}. The black hole is known to power a jet closely aligned along the line of sight, as indicated by the superluminal motions observed in the jet \cite{Orosz:2001dc}. The distance and sky location of the system, at Galactic longitude and latitude $l=6.8^\circ$ and $b=-4.8^\circ$, suggest that it belongs to the Bar of the Milky Way. A chart locating the source is shown in Fig. \ref{fig:cartoon}, which also outlines the geometry of the Galactic Bar based on Gaia measurements \cite{2019A&A...628A..94A}. 

The source appears very extended in \gr s, with a jet-like feature extending beyond $\sim 1^\circ-2^\circ$ from the source \cite{hawc,lhaaso} in the direction shown by the red line in Fig. \ref{fig:cartoon}. This jet-like feature is surprising because of the two-sided appearance and considerable extension of the jet-like structure that suggests its large inclination angle with respect to the line of sight \cite{hawc}. The difference between the low- and high-energy jet directions can hardly be attributed to variations of the black hole spin axis because the   \gr\ jet structure appears straight on $\sim 100-200$~pc scales, suggesting that the \gr\ jet direction has been rather stable in the past.



\section{A trace of a Galactic magnetic field line?}

One way to explain the misalignment of low- and high-energy jets is to assume that the \gr\ jet-like structure is not directly related to the jet produced by the black hole in the binary system. It is well possible that the black-hole jet loses its stability and terminates close to the source, and high-energy particles contained in the jet are released into the interstellar medium. We propose that the high-energy jet-like features are produced by particles streaming along the ordered magnetic field lines into the interstellar medium. This would readily explain the difference in the direction of the low- and high-energy jets: the low-energy jet is pointing along the spin axis of the black hole powering the microquasar, while the high-energy extended emission traces the propagation of particles escaping from the source through the GMF. 

If so, the properties of the extended \gr\ emission provide valuable information on the magnetic field in the Galactic Bar, about which there is little direct information. The jet-like appearance of the source is consistent with the possibility of strongly anisotropic diffusion taking place in the presence of a regular magnetic field with a strength larger than the strength of the turbulent field component~\cite{Giacinti:2017dgt}. The direction of the \gr\ jet shown by the red line in Fig.~\ref{fig:cartoon} can be interpreted as the direction of the GMF at the source location, projected on the plane of the sky. A more detailed study of the source morphology, such as the energy dependence of the length and width of the extended emission, can potentially be used to measure the diffusion coefficients parallel and perpendicular to the magnetic field and their energy dependence. 

\section{\label{sec:Visibility}Expected neutrino flux from hadronic interactions} 

A study of the diffusion coefficient and its energy dependence would have to rely on measuring the energies of particles emitting \gr s in the TeV-PeV energy range. Currently, two possibilities can be considered \cite{hawc}: inverse Compton emission from electrons or pion decay emission from interactions of high-energy protons. In the former case, the energies of electrons are comparable to those of the observed \gr s (as the inverse Compton scattering proceeds in the Klein-Nishina regime). In the latter case, the energies of protons reach above 10~PeV, more than an order of magnitude higher than the energies of the observed \gr s.

\begin{figure}
\centering
    \includegraphics[scale=0.4]{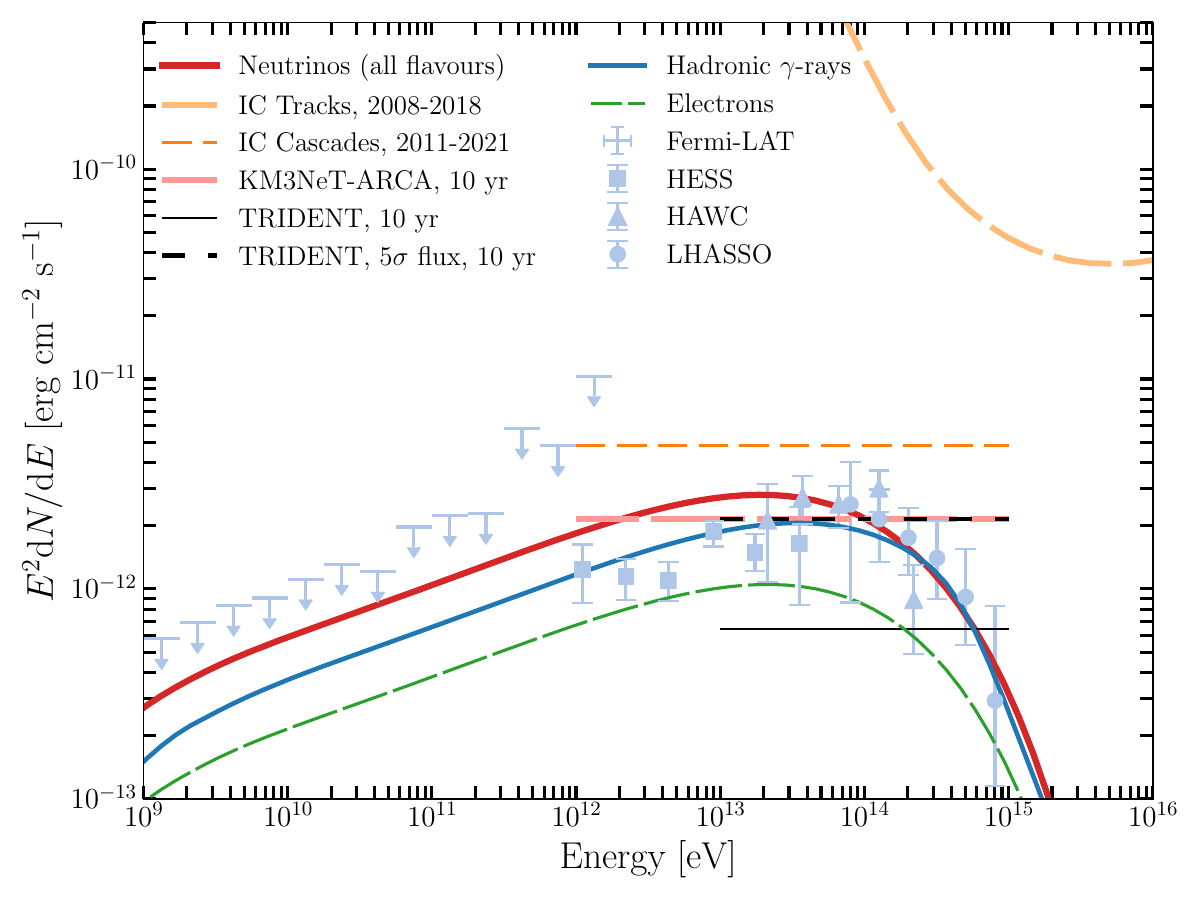}
\caption{\label{fig:uhe_sed}The high-energy spectral energy distribution of \v4641. The $68\%$ CL all-flavour neutrino upper limit of \ic (IC) and $90\%$ CL sensitivities of upcoming neutrino telescopes are shown under the assumption of an $E^{-2}$ power-law neutrino spectrum, together with \gr\ spectra measured by HESS~\cite{hess}, HAWC~\cite{hawc}, and LHASSO~\cite{lhaaso}, and Fermi-LAT upper limits which we obtained (see Supplementary Material). We show the sensitivity of \km-ARCA~\cite{KM3NeT:2018wnd} and the $5\sigma$ discovery flux of the proposed TRIDENT neutrino telescope~\cite{Ye:2022vbk}. The model lines show a fit to the \gr\ spectrum of \v4641 (blue line), the accompanying neutrino emission (red line), and the accompanying electron and positron emission (green line) under the assumption of a purely hadronic origin, calculated with AAFrag model (see text for details).}
\end{figure}

In the case of \v4641, it is challenging, but not impossible, to directly determine the type of particles escaping from the source. One expects to observe an accompanying neutrino flux if these particles are protons or nuclei. The source is within the sensitivity reach of neutrino telescopes, as illustrated by Figure \ref{fig:uhe_sed}. This figure summarises the measurements of the \gr\ flux of the source together with a model of  \gr\ and neutrino fluxes produced in proton-proton interactions simulated with AAfragpy~\cite{Koldobskiy_2021}. 
The assumed proton spectrum follows a power-law ${\rm d}N/{\rm d}E \propto E^{\alpha}$ with index $\alpha = -1.8$ and exhibits an exponential cutoff at 5 PeV. The all-flavour neutrino flux level is expected to reach $\sim 2\times 10^{-12}$~erg/(cm$^2$s) in the 10-100~TeV band.

This flux level is clearly out of IceCube's reach in its muon track detection channel because it is situated in the Southern Hemisphere. Fig.~\ref{fig:uhe_sed} shows the upper limit on the neutrino flux derived from 
the publicly available 10-year track dataset (2008 - 2018)~\cite{IceCube:2021xar} using the Multi-Messenger Online Data Analysis (MMODA) service \footnote{https://mmoda.io} powered by the  SkyLLH~\cite{IceCube:2023ihk} analysis software. In our calculation, we have taken into account that the angular resolution of IceCube, $\theta\simeq 0.2^\circ$ is better than the source size, and in fact, the two-degree-long jet-like feature would be resolved in neutrinos into $\simeq 5$ sources so that the flux of each source is $\simeq 1/5$ of the total. In such a situation, the sensitivity is degraded by a factor of $\sqrt{5}$.  

The sensitivity of IceCube in the cascade detection channel is much better in the Southern hemisphere. Comparing the point source sensitivity of this channel reported in Ref.~\cite{IceCube:2023ame} for the $E^{-2}$ type power-law spectra, $F_{fl}\simeq 10^{-12}$~TeV/(cm$^2$~s), for the per-flavour neutrino flux at the source declination, one finds that the upper limit on the all-flavour source flux is $3F_{fl}\simeq 5\times 10^{-12}$~erg/(cm$^2$~s), somewhat above the expected source flux level. Note that in the case of the cascade event channel, no correction for the extended nature of the source is needed because the angular resolution in this channel is worse than the source size. 
It is interesting to note that \v4641 is located within one of the excess regions in the significance map of the IceCube cascade data analysis of Ref.~\cite{IceCube:2023ame} so that the source may be contributing to this excess.  

Contrary to IceCube, the \km-ARCA detector that is currently under construction will be able to observe the source in both muon track and cascade detection channels. To estimate the detectability of \Sgr, we use the projected track sensitivity and discovery potential of \km-ARCA~\cite{KM3NeT:2018wnd}. We show this estimate in Figure~\ref{fig:uhe_sed} for the all-flavour neutrino flux. Similar to the IceCube track channel, we introduce a correction factor $\sqrt{5}$ to the sensitivity to account for the extended nature of the source. Clearly, detecting the source will also be challenging for \km: the flux barely reaches the sensitivity level. We conclude that a multi-km$^3$ size detector is needed for a sensible study of spectral and morphological characteristics of the neutrino source. For example, we show in Fig.~\ref{fig:uhe_sed} the sensitivity reach of the TRIDENT multi~km$^3$ detector proposed for installation in the South China Sea~\cite{Ye:2022vbk}. With such a telescope, the source will be detected at the $\sim 5\sigma$ significance level.

\section{X-ray synchrotron emission from  electrons} \label{sec:synchrotron}

Even if the observed \gr\ emission is from high-energy proton interactions, secondary electrons from the pion decays should still be present in the emission region. Emission from these secondary or primary electrons provides a possibility for an additional measurement that is directly useful in studying the GMF and high-energy particle propagation through it.  In the case of secondary electrons from high-energy proton interactions, their injection spectrum is expected to be close to the \gr\ and neutrino spectrum, i.e. $dN/dE\sim E^{-\Gamma},\ \Gamma\simeq -1.8$, in the energy range up to $E_{max}\gtrsim 100$~TeV, as shown in Fig.~\ref{fig:uhe_sed}.

Both primary and secondary electrons from proton interactions cool through synchrotron and inverse Compton scattering that softens their spectrum to $dN/dE\sim E^{-(\Gamma+1)}$. The inverse Compton emission contributes to the overall \gr\ flux. Its potential presence complicates the analysis of the \gr\ data because it is difficult to disentangle the pion decay and inverse Compton components based on \gr\ measurements alone. 

\begin{figure}
\includegraphics[width=\columnwidth]{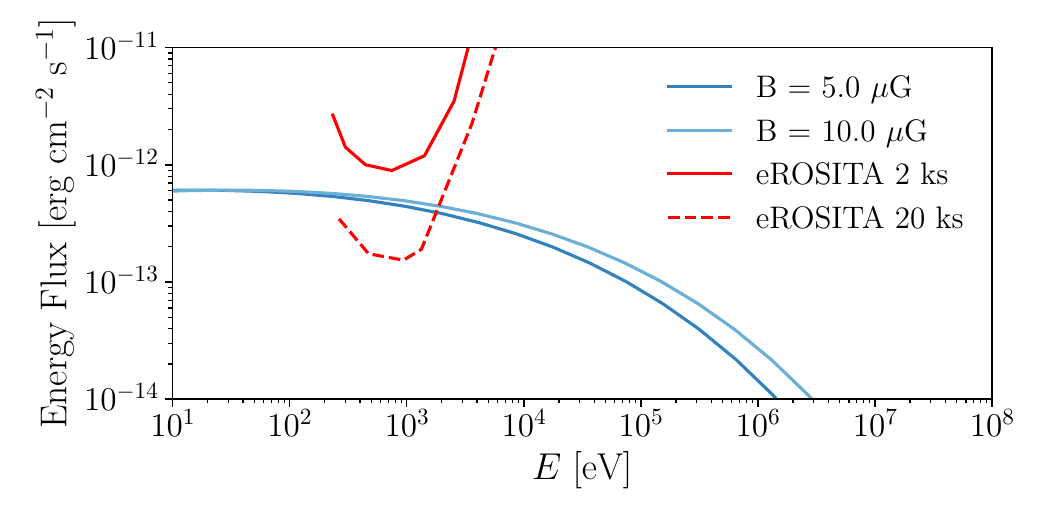}
\caption{Synchrotron emission from hadronic electrons that do not escape the 200-pc-size jet of \v4641. The sensitivity of a 2 or 20 kilosecond exposure of eROSITA accounting for the extended size of the source is shown for comparison\label{fig:synchrotron}.}
\end{figure}

Synchrotron emission from electrons is emitted in the X-ray band, 
\begin{equation}
    E_s=5\left[\frac{E_e}{100\mbox{ TeV}}\right]^2\left[\frac{B}{10\ \mu\mbox{G}}\right]\mbox{ keV,}
\end{equation}
with an $\sim E^{-(\Gamma/2+1)}\sim E^{-3}$ type spectrum below this characteristic energy. Fig.~\ref{fig:synchrotron} shows the spectrum of the synchrotron emission from the secondary electrons produced in interactions of protons within the same model as used for the calculation of Fig.~\ref{fig:uhe_sed}.

The total power in the secondary synchrotron flux is determined by the power injection from proton-proton interactions. It is largely independent of the magnetic field strength as long as electrons are efficiently cooled in this magnetic field. The interstellar radiation field in the Galactic Bar region has density $U_{rad}\sim 10$~eV/cm$^3$ in the micrometer frequency range \cite{Moskalenko:2005ng}. The 100~TeV electrons are mostly scattered on longer-wavelength far-infrared photons with one order-of-magnitude smaller density in the $U_{rad}\sim 1$~eV/cm$^3$ range. The synchrotron loss dominates over the inverse Compton loss as long as the energy density of the magnetic field $U_B\simeq 3[B/10\ \mu$G$]^2$~eV/cm$^3$ is larger than the energy density of the radiation. 

Measurements of the synchrotron flux can test the nature of the \gr\ emission (hadronic or leptonic) even in the absence of detection of the source in neutrinos. If the \gr\ source is leptonic, the X-ray synchrotron flux level can vary widely, depending on the magnetic field strength. To the contrary, the synchrotron flux from secondary electrons in a hadronic source is fixed by the observed \gr\ flux. 

Detection of the source in X-rays also poses a particular challenge because of the large size of the source.  The source flux level, $F_s\lesssim 10^{-12}$~erg/(cm$^2$s) is below the sensitivity limit of the ROSAT all sky survey \cite{2016A&A...588A.103B}, but may be within the reach of eROSITA \cite{eROSITA:2020emt}. 
Fig.~\ref{fig:synchrotron} shows that the synchrotron emission is detectable with a $\sim 20$~kilosecond exposure of eROSITA for magnetic field strengths in the 5-10~$\mu$G range.



Observations of the synchrotron flux from the source may also be used to measure the GMF strength if the measurements can be extended in the hard X-ray band, where one expects a magnetic-field-dependent high-energy cut-o ff in the source spectrum. 

\section{Discussion and conclusions \label{discuss}}

We have discussed a possible interpretation of the remarkable feature of the \gr\ signal from the microquasar \v4641~\cite{hawc,hess,lhaaso} that possesses a jet-like extension with length exceeding 100 pc at TeV energy~\cite{hess,hawc} and up to 200 pc at 100 TeV~\cite{hawc,lhaaso}, misaligned with the radio jet emitted by the black hole in the system. Within our interpretation, the extended feature traces high-energy particles spreading along the ordered GMF field line and interacting with the interstellar medium.

The  \gr\ spectral characteristics of the source, shown in Fig.~\ref{fig:uhe_sed}, can be well modelled within the hadronic model of the source activity. The location of \v4641 outside of the Galactic Plane would normally leave high-energy protons from the source without targets for proton-proton interactions. However, \v4641 is located in the  Galactic Bar, see Fig.~\ref{fig:cartoon}, where the density of the material may be high enough. We discussed the prospects for the definitive test of the hadronic nature of the high-energy emission via the detection of neutrinos. \v4641 is located in the Southern Hemisphere, where detecting muon neutrinos with IceCube is difficult. Nevertheless, the source will be marginally detectable with KM3NeT and detectable at a high significance level with  multi-km3 size detectors  TRIDENT~\cite{Ye:2022vbk} or HUNT \cite{Huang:2023mzt}. We have also discussed a complementary test of the hadronic nature of the source via the detection of X-ray synchrotron emission from secondary electrons originating from proton interactions.

The jet-like structure observed by HAWC~\cite{hawc} appears to have a length that increases from 100~pc at TeV energy up to 200 pc (projected on the sky) at 100~TeV, as also observed by LHAASO \cite{lhaaso}.  Within our model, such energy dependence is expected in the hadronic model of \gr\ emission because the cosmic ray scattering length is expected to grow with energy. Moreover, modelling of Galactic cosmic ray propagation shows that it may reach the characteristic correlation length distance scale of the turbulent magnetic field in the PeV energy range where the "knee" of the cosmic ray spectrum is found~\cite{Giacinti:2017dgt}. In principle, such a transition should be observable as a sharp increase in the length of the jet-like feature at the energy. 
The energy of the transition depends on the coherence length of the magnetic field.

Most importantly, we have shown that a combination of \gr\, neutrino and X-ray observations of sources like the \v4641\ jet can provide a tool for the measurement of the GMF characteristics (direction of its ordered component, relative strength of the ordered and turbulent components, the overall field strength, coherence length) at the source location. In the specific case of the \v4641\ system, observational data may provide information on the magnetic field in the Galactic Bar, indicating a moderate level of the turbulent component of the GMF in the Bar compared to the ordered component. This is valuable information because the geometry of the GMF in the central part of the Milky Way is highly uncertain. The existing GMF models make consistent predictions in the field in the halo of Milky Way \cite{2012ApJ...757...14J,Unger:2023lob,Korochkin:2024yit}, but they are currently not able to reliably constrain the magnetic field in the complicated region of the central Galaxy.

\begin{acknowledgments}
\noindent We thank Karri Koljonen for useful information about X-ray observations of \v4641. The work of D.S. and A.N. has been supported in part by the French National Research Agency (ANR) grant ANR-24-CE31-4686.
\end{acknowledgments}




\bibliography{refs}

\section{Appendix}

\subsection{Fermi-LAT data analysis}

Fig. \ref{fig:uhe_sed} shows the Fermi/LAT upper limit on the source flux. We have extracted this upper limit from publicly available Fermi/LAT data for the period between 2008 and 2024 provided by Fermi Science Support Center\footnote{https://fermi.gsfc.nasa.gov/ssc/}. We have filtered the data to retain the events of We filter the events of P8R3\_SOURCEVETO\_V3 type using the {\tt gtselect-gtmktime} 
tool chain to retain only the highest quality events most likely expected to be \gr s. We defined a rectangular source region of width 0.5 degrees in Right Ascension, and two degrees in Declination, centered on the location of \v4641 and estimated the background from two regions of identical geometry shifted by $\pm 2.5^\circ$ in Galactic longitude.  No significant source was detected, and we extracted flux upper limits shown in Fig.~\ref{fig:uhe_sed}. 

\subsection{Synchrotron emission}
\label{subsec:synchrotron}
In the main text, we discussed the synchrotron emission from the hadronic \gr\ electrons coproduced with \gr s and neutrinos under the assumption of a purely hadronic origin of the \gr\ emission shown in Fig.~\ref{fig:uhe_sed}. The electrons radiate synchrotron photons in the presence of the magnetic field of the Milky Way. To estimate the synchrotron emission, we first determined the steady-state electron spectrum in the extended $E> 100$ TeV \gr\ emitting region. To do so, we used the transport equation
\begin{equation}
\frac{{\rm}d}{{\rm d}E_e} \left[\frac{{\rm d}E_e}{{\rm d}t} N(E_e) \right]= \dot{N}(E_e). 
\end{equation}
\noindent where $N(E_e)$ is the number spectrum of electrons and $\dot{N}(E_e)$ is the source term. 
This equation has the analytical solution~\cite{Dermer:2009zz}
\begin{equation}
    N(E_e) = \left|\frac{{\rm d}E_e}{{\rm d}t}\right|^{-1}\int_{E_e}^{\infty}{\rm d}E_e' N(E_e)'
\end{equation}
which we numerically integrate. 
The energy lost by these electrons to synchrotron radiation is calculated as 
\begin{equation}
    \frac{{\rm d}E_e}{{\rm d}t} = -\frac{4}{3} \sigma_{\rm T}c\gamma_e^2 U_B. 
\end{equation}
where $\gamma_e$ is the Lorentz factor of electrons of energy $E_e$, $\sigma_{\rm T}$ is the Thomson crosssection and $U_B = B^2/8\pi$ is the energy density of the magnetic field.  We then calculate the synchrotron power radiated per unit photon energy following Eq. 9 of Ref.~\cite{1988ApJ...334L...5G}. Fig.~\ref{fig:synchrotron} shows the synchrotron emission from hadronic electrons in a magnetic field with strength ranging from 3-10$\mu$G. We also show the sensitivity of eROSITA to such emission, taken from Ref.~\cite{erosita}. 
We account for the reduced sensitivity of eROSITA due to the large extension of the source. The point-spread function of eROSITA is $R\approx26''$~\cite{Brunner_2022} while the total area of the $E>100$~TeV image of \v4641 is approximately 1~degree$^2$. We therefore multiply the nominal sensitivity by $\sqrt{1~{\rm degree}^2/(\pi R^2)} \sim 100$. This is shown in Fig.~\ref{fig:synchrotron} for a two-kilosecond and a 20-kilosecond exposure.


\end{document}